\documentclass[aps,prb,twocolumn,showpacs]{revtex4}

\usepackage{hyperref,latexsym}
\usepackage{graphicx,color,pstricks}
\usepackage{epsfig,epsf,bm}

\begin{document}
\title{Quantum dynamics of tunneling between ferromagnets}
\author{Yu-Li Lee}
\affiliation{Physics Department, National Changhua University of
Education, Changhua, Taiwan, R.O.C.}
\email{yllee@cc.ncue.edu.tw}
\author{Yu-Wen Lee}
\affiliation{Physics Department, Tunghai University, Taichung,
Taiwan, R.O.C.}
\email{ywlee@mail.thu.edu.tw}
\begin{abstract}
 We study the Josephson-like spin currents between two
 ferromagnetic metals by deriving the effective action of the
 junction. A DC spin Josephson current with the full O($3$)
 symmetry is obtained. We also show that a time-independent uniform
 magnetic field can serve as the source of the AC spin Josephson
 effect. That is, the spin current in a uniform magnetic field becomes
 a periodic function of the time with the period proportional to the
 inverse of the magnitude of the external magnetic field.
\end{abstract}
 \maketitle

\section{Introduction}

One of the striking phenomena about the superconductors is the
Josephson effect\cite{J} in the superconducting (SC) tunnel
junctions. The Josephson effect arises from the fact that the
phases of the SC order parameters of the two superconductors tend
to become uniform when they are coupled to each other. A natural
question associated with the Josephson effects is what happens
when two systems with another type of long range orders are weakly
coupled? We shall partly address this question by considering the
tunnel junction between two ferromagnets. This is the simplest
extension of the SC tunnel junctions because the underlying
symmetry behind the ferromagnets is O($3$), while the occurrence
of the SC long range order is a realization of the spontaneous
U($1$) symmetry breaking.

In recent years, the possibility of using the spin degrees of
freedom in the electronic devices, known as spintronics, receives
considerable attention and is a rapidly developing research topic.
In this field, the manipulation of the spin current is a subject
of extensive investigation. An interesting extreme case of a
finite spin current without charge currents has been investigated
by several groups.\cite{H,BTWH,WWG} Also, the spin transport
without dissipation in thin film ferromagnets was discussed
recently.\cite{KBD} In analogy with the SC junctions, a
Josephson-like spin current may occur in the ferromagnetic (FM)
tunnel junctions. Therefore, the study of the FM junctions is
intimately connected with the control of the spin transport.

Indeed, a DC Josephson-like spin current occurring in the FM
junctions was predicted recently.\cite{NB} However, in Ref.
\onlinecite{NB}, only the effects of the U($1$) phase, which
corresponds to a subgroup of the full O($3$) symmetry, were
explored. In the present paper, we will treat the bulk
ferromagnets within the framework of the Stoner
ferromagnetism\cite{St} and study this problem by taking the
effective action approach, parallel to the one in the
investigation of the SC junctions.\cite{AES} This approach has
several advantages. First, the O($3$) symmetry is manifestly
respected, and thus the effects of other degrees of freedom in
addition to the U($1$) phase considered in Ref. \onlinecite{NB}
can be revealed. Next, the roles of the quasiparticles and
collective modes, especially the dissipation due to the
quasiparticle tunneling, are explicitly disentangled. Moreover, a
renormalization group analysis about the relevancy of the
tunneling action can be performed. Finally, with the help of the
effective action, the calculations of the spin current and its
correlation functions become straightforward. Our main results are
as follows: (i) We derive an effective action of the FM tunnel
junction. (ii) The DC spin Josephson current and current noise are
obtained. The latter exhibits the Johnson-Nyquist form at low
temperature. (iii) We show that a {\it time-independent uniform
magnetic field} can serve as the source of the AC spin Josephson
effect. That is, the spin current in a uniform magnetic field
becomes a periodic function of the time with the period
proportional to the inverse of the magnitude of the external
magnetic field.

The rest of the paper is organized as follows: The derivation of
the effective action of the tunnel junction is given at Sec.
\ref{effact}. We calculate the DC and AC spin Josephson currents
as well as the current noise in Sec. \ref{sje}. Sec. \ref{rga} is
devoted to the perturbative renormalization group (RG) analysis of
the effective action, where the possible role of the quasiparticle
tunneling played in the dissipation is examined. Finally, we
discuss our results and experimental implications in the last
section.

\section{Effective action}
\label{effact}

We start with the following action
\begin{equation}
 S=\int^{\beta}_0 \! \! d\tau \left[\int \! \! d^dx~
  ({\mathcal L}_l+{\mathcal L}_r)+L_T\right] \ ,
  \label{jact1}
\end{equation}
where $d$ is the spatial dimensions. Here ${\mathcal L}_l$ and
${\mathcal L}_r$ describe the ferromagnetic metals on the left and
right of the junction. We adopt the Stoner model for the itinerant
ferromagnetism.\cite{St} The corresponding Lagrangian is given by
\begin{equation}
 {\mathcal L}_l=\psi^{\dagger}_l\left[\partial_{\tau}-\frac{\nabla^2}
               {2m_l}-\mu -\Delta_l\bm{\Omega}_l\cdot \sigma_z\right]
               \psi_l \ , \label{jlag1}
\end{equation}
and a similar expression with $l\rightarrow r$. Here
$\psi_{l(r)\alpha}$ is the electron operator with spin $\alpha$ on
the left (right) side of the junction, $\bm{\Omega}_{l(r)}$ is a
unit vector, and $\Delta_{l(r)}>0$ is proportional to the
magnitude of the bulk magnetization on the left (right) of the
junction. In Eq. (\ref{jlag1}), the term proportional to
$\Delta^2_l$ is not written down explicitly because it is only
related to the determination of $\Delta_l$ and is not important
for the following discussions. We will treat $\Delta_{l(r)}$ as a
given number which is determined, for example, by the mean-field
theory, and consider the fluctuations of $\bm{\Omega}_{l(r)}$
only. $L_T$ is the tunneling Lagrangian which is of the form
\begin{equation}
 L_T=\int_{\bm{x}\in l\atop \bm{x}^{\prime}\in r} \! \! d^dx
    d^dx^{\prime}\left[T(\bm{x},\bm{x}^{\prime})\psi^{\dagger}_l
    (\bm{x})\psi_r(\bm{x}^{\prime})+{\mathrm H.c.}\right] \ .
    \label{jlag2}
\end{equation}
We shall follow a procedure similar to Ref. \onlinecite{AES} to
derive the effective action of the FM tunnel junction.

To proceed, we make the gauge transformation
\begin{eqnarray}
 \tilde{\psi}_{l(r)}(x) &=& g_{l(r)}^{\dagger}(x)\psi_{l(r)}(x)
       \ , \nonumber \\
 \tilde{\psi}_{l(r)}^{\dagger}(x) &=& \psi_{l(r)}^{\dagger}(x)
       g_{l(r)}(x) \ , \label{jgauge1}
\end{eqnarray}
where $x_{\mu}=(\tau ,\bm{x})$ and $g_{l(r)}$ is an SU($2$) matrix
which satisfies the relation
\begin{equation}
 g_{l(r)}\sigma_zg_{l(r)}^{\dagger}=\bm{\Omega}_{l(r)}\cdot \bm{\sigma}
         \ .\label{su21}
\end{equation}
Then, ${\mathcal L}_{l(r)}$ and $L_T$ become
\begin{widetext}
\begin{eqnarray}
 {\mathcal L}_{l(r)} &=& \tilde{\psi}^{\dagger}_{l(r)}\left[
                     \partial_{\tau}-\frac{\nabla^2}{2m_{l(r)}}
                     -\mu -\Delta_{l(r)}\sigma_z\right]
                     \tilde{\psi}_{l(r)}+\rho_{l(r)\alpha \beta}
                     \left(g^{\dagger}_{l(r)}\partial_{\tau}
                     g_{l(r)}\right)_{\alpha \beta} \nonumber \\
                     & & -i\bm{J}_{l(r)\alpha \beta}\cdot \left(
                     g^{\dagger}_{l(r)}\bm{\nabla}g_{l(r)}
                     \right)_{\alpha \beta}-\frac{1}{2m_{l(r)}}~
                     \rho_{l(r)\alpha \beta}\left[\left(
                     g^{\dagger}_{l(r)}\bm{\nabla}g_{l(r)}\right)^2
                     \right]_{\alpha \beta}\ , \label{jlag3} \\
 L_T &=& \int_{\bm{x}\in l\atop \bm{x}^{\prime}\in r} \! \! d^dx
     d^dx^{\prime}\left[T(\bm{x},\bm{x}^{\prime})
     \tilde{\psi}^{\dagger}_l(\bm{x})g^{\dagger}_l(\bm{x})g_r
     (\bm{x}^{\prime})\tilde{\psi}_r(\bm{x}^{\prime})+{\mathrm H.c.}
     \right] \ , \label{jlag4}
\end{eqnarray}
where $\rho_{l(r)\alpha
\beta}=\tilde{\psi}^{\dagger}_{l(r)\alpha}\tilde{\psi}_{l(r)\beta}$
and $\bm{J}_{l(r)\alpha
\beta}=\frac{1}{2m_{l(r)}}\left[\tilde{\psi}^{\dagger}_{l(r)\alpha}
(-i\bm{\nabla})\tilde{\psi}_{l(r)\beta}+i\bm{\nabla}
\tilde{\psi}^{\dagger}_{l(r)\alpha}\tilde{\psi}_{l(r)\beta}\right]$.
By integrating out the fermion fields, the partition function is
written as
\begin{eqnarray*}
 {\mathcal Z}=\int \! \! D[g^{\dagger}_l]D[g_l]D[g^{\dagger}_r]D[g_r]
             \exp{\left\{-{\mathcal A}\left[g,g^{\dagger}\right]
             \right\}} \ ,
\end{eqnarray*}
where
\begin{equation}
 {\mathcal A}\left[g,g^{\dagger}\right]=-{\mathrm tr}\left[
             \ln{\left(\underline{{\mathcal G}}^{-1}\right)}\right]
             \ . \label{jact2}
\end{equation}
Here we have introduced a four-component fermion space by adding
the spinor spaces of the left and right ferromagnets. In
particular,
\begin{eqnarray}
 \underline{{\mathcal G}}^{-1} &=& \left(\begin{array}{cc}
           \hat{{\mathcal G}}_l^{-1} & -\hat{{\mathcal T}} \\
           -\hat{{\mathcal T}}^{\dagger} & \hat{{\mathcal G}}_r^{-1}
           \end{array}\right) \ , \label{jgf1} \\
 \hat{{\mathcal T}} &=& T(\bm{x},\bm{x}^{\prime})g^{\dagger}_l(x)g_r
                    (x^{\prime})\delta (\tau -\tau^{\prime}) \ ,
                    \label{jgf2}
\end{eqnarray}
and
\begin{equation}
 \hat{{\mathcal G}}_{l(r)}^{-1}=-\left\{\partial_{\tau}
                  +g^{\dagger}_{l(r)}\partial_{\tau}g_{l(r)}-\frac{1}
                  {2m_{l(r)}}\left[\bm{\nabla}+g^{\dagger}_{l(r)}
                  \bm{\nabla}g_{l(r)}\right]^2-\mu -\Delta_{l(r)}
                  \sigma_z\right]\delta (x-x^{\prime}) \ .
                  \label{jgf3}
\end{equation}
Note that we indicate matrices in the spinor space of one
ferromagnet by carets, and matrices acting on the four-component
fermion space by underlines.

Now we expand the right-hand side of Eq. (\ref{jact2}) in powers
of the tunneling matrix elements, namely the off-diagonal parts of
Eq. (\ref{jgf1}). Keeping the lowest nonvanishing terms, we obtain
\begin{equation}
 {\mathcal A}={\mathcal A}_l+{\mathcal A}_r+{\mathrm tr}
             \left[\hat{{\mathcal G}}_l~\hat{T}~\hat{{\mathcal G}}_r~
             \hat{T}^{\dagger}\right] \ , \label{jact3}
\end{equation}
where ${\mathcal A}_{l(r)}$ is the bulk action of the left (right)
ferromagnet. For simplicity, we shall consider the point-like
junction, i.e. $T(\bm{x},\bm{x}^{\prime})=\widetilde{T}\delta
(\bm{x})\delta (\bm{x}^{\prime})$. In terms of the parametrization
$g_l(x)=h_l(x)g_{0l}$ and a similar expression with $l\rightarrow
r$, where $g_{0l(r)}$ satisfies the relation
$g_{0l(r)}\sigma_zg_{0l(r)}^{\dagger}=\bm{n}_{l(r)}\cdot
\bm{\sigma}$ and $\bm{n}_{l(r)}$ is a unit vector along the
direction of the magnetization in the left (right) ferromagnet, we
can expand ${\mathcal A}_{l(r)}$ in powers of
$h_{l(r)}^{\dagger}\partial_{\mu}h_{l(r)}$. (Here $g_{l(r)}$ or
$\bm{\Omega}_{l(r)}$ are decomposed into two parts: $g_{0l(r)}$ or
$\bm{n}_{l(r)}$, which gives the direction of the magnetization in
the bulk, is fixed and the matrix field $h_{l(r)}$ describes the
quantum (spin-wave) fluctuations.) Keeping the lowest nonvanishing
term and integrating out the degrees of freedom away from the
position of the junction, the resulting action is given by
\begin{equation}
 {\mathcal A}_l={\mathcal A}_l^0+M_l\int^{\beta}_0 \! \! d\tau ~
               \bm{n}_l\cdot {\mathrm tr}\left[\bm{\sigma}
               h_l^{\dagger}\partial_{\tau}h_l\right] \ ,
               \label{lact1}
\end{equation}
and the similar expression with $l\rightarrow r$. Here $M_{l(r)}$
is the magnetization per volume of the left (right) ferromagnet.
The effective action of the ferromagnetic junction is given by the
last term in Eq. (\ref{jact3}), the last term in Eq.
(\ref{lact1}), and a similar term with $l\rightarrow r$. Before
examining the third term in Eq. (\ref{jact3}), two points should
be mentioned. First, to arrive at Eq. (\ref{lact1}), the
interactions between the spin waves are neglected. Second, the
bulk action of the FM tunnel junction starts with the first-order
time derivative due to the Berry phase of quantum spins, whereas
for the SC tunnel junction, the bulk action starts with the
second-order time derivative.

Working out the third term in Eq. (\ref{jact3}), we find that it
is given by
\begin{equation}
 {\mathcal A}_T=|\widetilde{T}|^2\int^{\beta}_0 \! \! d\tau_1d\tau_2
               \! \! \int \! \! \frac{d^dp_1}{(2\pi)^d}\frac{d^dp_2}
               {(2\pi)^d}~{\mathrm tr}\left[h^{\dagger}_r(\tau_1)h_l
               (\tau_1)\hat{D}_l(\tau_1-\tau_2,\bm{p}_1)h^{\dagger}_l
               (\tau_2)h_r(\tau_2)\hat{D}_r(\tau_2-\tau_1,\bm{p}_2)
               \right] \ , \label{tact1}
\end{equation}
where
\begin{equation}
 \hat{D}_l(\tau ,\bm{p})=\int \! \! d^dx~e^{-i\bm{p}\cdot \bm{x}}g_{0l}
                        \hat{{\mathcal G}}_l(\tau ,\bm{x})g^{\dagger}_{0l}
                        =G_l(\tau ,\bm{p})\sigma_0-F_l(\tau ,\bm{p})
                        \bm{n}_l\cdot \bm{\sigma} \ , \label{gf1}
\end{equation}
with
\begin{eqnarray}
 G_l(\tau ,\bm{p}) &=& \frac{1}{\beta}\sum_ne^{-i\omega_n\tau}\frac{
                   i\omega_n-\epsilon_{l\bm{p}}+\mu}{\left(i\omega_n
                   -\epsilon_{l\bm{p}}+\mu \right)^2-\Delta_l^2} \ ,
                   \nonumber \\
 F_l(\tau ,\bm{p}) &=& \frac{1}{\beta}\sum_ne^{-i\omega_n\tau}
                   \frac{\Delta_l}{\left(i\omega_n-\epsilon_{l\bm{p}}
                   +\mu \right)^2-\Delta_l^2} \ , \label{gf2}
\end{eqnarray}
and a similar expression with $l\rightarrow r$. In the above,
$\sigma_0$ is the $2\times 2$ identity matrix and
$\omega_n=(2n+1)\pi T$. With the help of Eq. (\ref{gf1}), one may
find that ${\mathcal A}_T$ is composed of four terms,
\begin{eqnarray}
 {\mathcal A}_T &=& \int^{\beta}_0 \! \! d\tau_1d\tau_2\left\{
                {\mathcal I}_1(\tau_1-\tau_2){\mathrm tr}\left[
                {\mathcal M}^{\dagger}(\tau_1){\mathcal M}(\tau_2)
                \right]+{\mathcal I}_2(\tau_1-\tau_2){\mathrm tr}
                \left[(\bm{n}_r\cdot \bm{\sigma}){\mathcal M}^{\dagger}
                (\tau_1)(\bm{n}_l\cdot \bm{\sigma}){\mathcal M}(\tau_2)
                \right]\right. \nonumber \\
                & & \left.+{\mathcal I}_3(\tau_1-\tau_2){\mathrm tr}
                \left[{\mathcal M}^{\dagger}(\tau_1)(\bm{n}_l\cdot
                \bm{\sigma}){\mathcal M}(\tau_2)\right]+{\mathcal I}_4
                (\tau_1-\tau_2){\mathrm tr}\left[(\bm{n}_r\cdot \bm{\sigma})
                {\mathcal M}^{\dagger}(\tau_1){\mathcal M}(\tau_2)\right]
                \right\} \ , \label{tact2}
\end{eqnarray}
where ${\mathcal M}(\tau)=h^{\dagger}_l(\tau)h_r(\tau)$ and
\begin{eqnarray*}
 {\mathcal I}_1(\tau) &=& |\widetilde{T}|^2\int \! \! \frac{d^dp_1}
                      {(2\pi)^d}\frac{d^dp_2}{(2\pi)^d}~G_l
                      (\tau ,\bm{p}_1)G_r(-\tau ,\bm{p}_2) \ , \\
 {\mathcal I}_2(\tau) &=& |\widetilde{T}|^2\int \! \! \frac{d^dp_1}
                      {(2\pi)^d}\frac{d^dp_2}{(2\pi)^d}~F_l
                      (\tau ,\bm{p}_1)F_r(-\tau ,\bm{p}_2) \ , \\
 {\mathcal I}_3(\tau) &=& -|\widetilde{T}|^2\int \! \! \frac{d^dp_1}
                      {(2\pi)^d}\frac{d^dp_2}{(2\pi)^d}~F_l
                      (\tau ,\bm{p}_1)G_r(-\tau ,\bm{p}_2) \ , \\
 {\mathcal I}_4(\tau) &=& -|\widetilde{T}|^2\int \! \! \frac{d^dp_1}
                      {(2\pi)^d}\frac{d^dp_2}{(2\pi)^d}~G_l
                      (\tau ,\bm{p}_1)F_r(-\tau ,\bm{p}_2) \ .
\end{eqnarray*}
\end{widetext}
For $\epsilon_F^{-1}\ll |\tau|\ll T^{-1}$, we have
\begin{equation}
 {\mathcal I}_i(\tau)\approx -\alpha_i\frac{\pi^2T^2}
            {\sin^2{(\pi T\tau)}} ~~~~i=1,\cdots ,4 \ ,
            \label{tact4}
\end{equation}
where
\begin{eqnarray*}
 \alpha_1 &=& \frac{1}{4}~N_l(0)N_r(0)|\widetilde{T}|^2 \ , \\
 \alpha_2 &=& n_l(0)n_r(0)|\widetilde{T}|^2 \ , \\
 \alpha_3 &=& \frac{1}{2}~n_l(0)N_r(0)|\widetilde{T}|^2 \ , \\
 \alpha_4 &=& \frac{1}{2}~N_l(0)n_r(0)|\widetilde{T}|^2 \ .
\end{eqnarray*}
In the above,
$N_{l(r)}(0)=N_{l(r)\uparrow}(0)+N_{l(r)\downarrow}(0)$ and
$n_{l(r)}(0)=[N_{l(r)\uparrow}(0)-N_{l(r)\downarrow}(0)]/2$ where
$N_{l(r)\sigma}(\epsilon)$ is the density of states for spin
$\sigma$ electrons in the left (right) ferromagnet and $\epsilon
=0$ denotes the Fermi surface.

There are two points which should be emphasized. First, the
quasiparticle tunneling results in the non-local terms in
${\mathcal A}_T$. Secondly, compared with the SC tunnel junctions,
there are no terms like $\cos{\phi}$, which corresponds to
${\mathrm tr}\left[{\mathcal M}^2\right]+{\mathrm H.c.}$ in the
present case. In the expansion in powers of the tunneling matrix
elements, such a term arises from the nonvanishing anomalous Green
functions of electrons. However, in the FM case, the anomalous
Green functions of electrons vanish. Consequently, the terms like
${\mathrm tr}\left[{\mathcal M}^n\right]+{\mathrm H.c.}$ with some
positive integer $n$ cannot appear at any order in the expansion
of the tunneling matrix elements. This is a crucial distinction
between the SC and FM tunnel junctions. In conclusion, we have
derived the effective action of a FM tunnel junction, which is
given by
\begin{equation}
 {\mathcal A}_{eff}=\int^{\beta}_0 \! \! d\tau \left\{M_l\bm{n}_l
                   \cdot {\mathrm tr}\left[\bm{\sigma}h_l^{\dagger}
                   \partial_{\tau}h_l\right]+(l\rightarrow r)\right\}
                   +{\mathcal A}_T \ . \label{jact4}
\end{equation}

\section{Spin Josephson effect}
\label{sje}

Now we are able to study the spin currents and noises with the
help of the effective action ${\mathcal A}_{eff}$ [Eq.
(\ref{jact4})]. The spin current operator is defined by
\begin{equation}
 \bm{I}=\frac{d\bm{S}_l}{dt}=-i\left[\bm{S}_l,H\right] \ ,
       \label{spinc1}
\end{equation}
where $\bm{S}_l$ is the spin operator in the left ferromagnet
which is defined by $\bm{S}_l=\int \! \!
d^dx~\frac{1}{2}~\psi^{\dagger}_l(\bm{x})\bm{\sigma}\psi_l(\bm{x})$.
(Note that $\bm{I}=\frac{d\bm{S}_l}{dt}=-\frac{d\bm{S}_r}{dt}$ due
to the the conservation of the total spin.) After a
straightforward algebra, one may find that
\begin{equation}
 \bm{I}=\int_{\bm{x}_1\in l \atop \bm{x}_2\in r} \! \! d^dx_1d^dx_2
       \! \! \left[T(\bm{x}_1,\bm{x}_2)\frac{1}{2i}\psi^{\dagger}_l
       (\bm{x}_1)\bm{\sigma}\psi_r(\bm{x}_2)+{\mathrm H.c.}\right]
       \! \ . \label{spinc2}
\end{equation}
Therefore, the spin current and its correlation functions can be
calculated in terms of the generating functional
\begin{equation}
 {\mathcal Z}[\bm{\eta}]=\int \! \! D[u]~\exp{\left[-S+\int^{\beta}_0
             \! \! d\tau ~\bm{\eta}(\tau)\cdot \bm{I}(\tau)\right]}
             \ , \label{spinpf1}
\end{equation}
where the integration measure is defined by
$D[u]=D[\psi^{\dagger}_{l}]D[\psi_l]D[\psi^{\dagger}_r]D[\psi_r]$,
the action $S$ is given by Eq. (\ref{jact1}), and $\bm{\eta}$ is a
real source field. Using Eq. (\ref{spinpf1}), the spin current and
its two-point correlation function are given by
\begin{eqnarray}
 \langle I_a(\tau)\rangle &=& \frac{1}{{\mathcal Z}[0]}\left.\frac{
            \delta {\mathcal Z}[\bm{\eta}]}{\delta \eta_a(\tau)}
            \right|_{\bm{\eta}=0} \ , \label{spinc3} \\
 \langle I_a(\tau_1)I_b(\tau_2)\rangle &=& \frac{1}{{\mathcal Z}[0]}
            \left.\frac{\delta^2{\mathcal Z}[\bm{\eta}]}
            {\delta \eta_b(\tau_2)\delta \eta_a(\tau_1)}
            \right|_{\bm{\eta}=0} \ . \label{noise1}
\end{eqnarray}
The calculation of the generating functional ${\mathcal
Z}[\bm{\eta}]$ is parallel to the derivation of the effective
action of the tunnel junction ${\mathcal A}_{eff}$. The effect of
the addition of the source term $\bm{\eta}(\tau)\cdot
\bm{I}(\tau)$ is to replace $T(\bm{x}_1,\bm{x}_2)$ by
$T(\bm{x}_1,\bm{x}_2)\left[1+\frac{i}{2}\bm{\sigma}\cdot
\bm{\eta}(\tau)\right]$. This amounts to replacing ${\mathcal
M}(\tau)$ in Eq. (\ref{tact2}) by ${\mathcal
M}(\tau)+\frac{i}{2}\bm{\eta}(\tau)\cdot
h^{\dagger}_l(\tau)\bm{\sigma}h_r(\tau)$.

\paragraph{DC Josephson effect}

We first compute the spin current. According to Eq.
(\ref{spinc3}), it is given by
\begin{equation}
  \langle \bm{I}(\tau)\rangle =\frac{1}{{\mathcal Z}[0]}\int \! \!
          D[u]~e^{-{\mathcal A}_{eff}}\bm{I}[h_l,h_r;\tau] \ ,
          \label{spinc4}
\end{equation}
where $D[u]=D[h^{\dagger}_l]D[h_l]D[h^{\dagger}_r]D[h_r]$, and
\begin{widetext}
\begin{eqnarray}
 \bm{I}[h_l,h_r;\tau] &=& -\frac{i}{2}\int^{\beta}_0 \! \!
       d\tau^{\prime}\left\{{\mathcal I}_1(\tau -\tau^{\prime})
       {\mathrm tr}\left[{\mathcal M}^{\dagger}(\tau^{\prime})
       h^{\dagger}_l(\tau)\bm{\sigma}h_r(\tau)\right]\right.
       \nonumber \\
       & & \left.+{\mathcal I}_2(\tau -\tau^{\prime}){\mathrm tr}
       \left[(\bm{n}_r\cdot \bm{\sigma}){\mathcal M}^{\dagger}
       (\tau^{\prime})(\bm{n}_l\cdot \bm{\sigma})h^{\dagger}_l
       (\tau)\bm{\sigma}h_r(\tau)\right]\right. \nonumber \\
       & & \left.+{\mathcal I}_3(\tau -\tau^{\prime}){\mathrm tr}
       \left[{\mathcal M}^{\dagger}(\tau^{\prime})(\bm{n}_l\cdot
       \bm{\sigma})h^{\dagger}_l(\tau)\bm{\sigma}h_r(\tau)\right]
       \right. \nonumber \\
       & & \left.+{\mathcal I}_4(\tau -\tau^{\prime}){\mathrm tr}
       \left[(\bm{n}_r\cdot \bm{\sigma}){\mathcal M}^{\dagger}
       (\tau^{\prime})h^{\dagger}_l(\tau)\bm{\sigma}h_r(\tau)
       \right]-{\mathrm H.c.}\right\} \ . \label{c1}
\end{eqnarray}
\end{widetext}
Within the semiclassical approximation where
$h_l(\tau)=\sigma_0=h_r(\tau)$, the spin current becomes
\begin{equation}
 \langle \bm{I}(\tau)\rangle =2(\bm{n}_r\times \bm{n}_l)
         \int^{\beta}_0 \! \! d\tau^{\prime}~{\mathcal I}_2
         (\tau -\tau^{\prime}) \ . \label{spinc6}
\end{equation}
Note that the contributions to the spin current arising from all
terms except the second one in $\bm{I}[h_l,h_r;\tau]$ [Eq.
(\ref{c1})] vanish. Using Eq. (\ref{tact4}), the integral in Eq.
(\ref{spinc6}) can be evaluated and the result is
\begin{equation}
 \langle \bm{I}\rangle =\pi T\tau_0\cot{(\pi T\tau_0)}I_0
        (\bm{n}_l\times \bm{n}_r)\rightarrow I_0(\bm{n}_l\times
        \bm{n}_r) \ , \label{spinc7}
\end{equation}
where $I_0=4n_l(0)n_r(0)|\widetilde{T}|^2/\tau_0$ is the critical
current and $\tau_0\sim \epsilon_F^{-1}$ is an IR cut-off. Eq.
(\ref{spinc7}), which is valid only at low temperature, is the
analogy of the DC Josephson effect in the SC tunnel junctions.
Note that $\langle \bm{I}\rangle =0$ when the directions of the
magnetizations in the left and right ferromagnets are parallel or
anti-parallel to each other. Moreover, the critical current is
proportional to the difference between the densities of states of
electrons with spin up and down at the Fermi surface. Therefore,
the existence of the spin current we obtained requires that the
metals on both sides of the junction must exhibit long range FM
orders simultaneously.

\paragraph{AC Josephson effect}

For the SC tunnel junctions, an applied DC bias will induce an AC
Josephson current. The effect of the DC bias in that case is to
make a time-dependent phase rotation or a U($1$) gauge
transformation on the SC order parameter. In the FM case, an
analogous AC spin Josephson effect may be induced by making a spin
SU($2$) gauge transformation on the magnetization. One of the way
to achieve this goal is to add uniform magnetic fields. Based on
this observation, we consider the effects of uniform magnetic
fields, which is described by the Zeeman term
\begin{eqnarray}
 H_Z &=& -\int_{\bm{x}\in l} \! \! d^dx~\bm{B}_l\cdot \frac{1}{2}~
     \psi^{\dagger}_l(\bm{x})\bm{\sigma}\psi_l(\bm{x}) \nonumber
     \\
     & & -\int_{\bm{x}\in r} \! \! d^dx~\bm{B}_r\cdot \frac{1}{2}~
     \psi^{\dagger}_r(\bm{x})\bm{\sigma}\psi_r(\bm{x}) \ ,
     \label{zee1}
\end{eqnarray}
where $\bm{B}_{l(r)}$ is the external magnetic field exerted on
the left (right) ferromagnet. (Here, for simplicity, the constant
$g\mu_B$ is absorbed into $\bm{B}_{l(r)}$ where $g$ is the
gyromagnetic ratio and $\mu_B$ is the Bohr magneton. Furthermore,
the orbital effects are neglected.) Eq. (\ref{zee1}) can be
eliminated by performing the gauge transformation in the real-time
formalism
\begin{eqnarray}
 \psi_{l(r)} &\rightarrow& \exp{\left\{\frac{i}{2}~t\bm{B}_{l(r)}\cdot \bm{\sigma}\right\}}
            \psi_{l(r)} \ , \nonumber \\
 \psi_{l(r)}^{\dagger} &\rightarrow& \psi_{l(r)}^{\dagger}
            \exp{\left\{-\frac{i}{2}~t\bm{B}_{l(r)}\cdot \bm{\sigma}\right\}}
            \ . \label{zee2}
\end{eqnarray}
In the imaginary-time formulation, $t$ and $\bm{B}_{l(r)}$ in Eq.
(\ref{zee2}) are replaced by $-i\tau$ and $i\bm{B}_{l(r)}$,
respectively. Under the gauge transformation (\ref{zee2}), the
tunneling matrix $\hat{{\mathcal T}}$ [Eq. (\ref{jgf2})] becomes
\begin{equation}
 \hat{{\mathcal T}}\rightarrow T(\bm{x},\bm{x}^{\prime})g^{\dagger}_l(x)
                   U^{\dagger}_l(\tau)U_r(\tau)g_r(x^{\prime})\delta
                   (\tau -\tau^{\prime}) \ , \label{tmatrix1}
\end{equation}
where $U_{l(r)}(\tau)=\exp{\left\{\frac{i}{2}~\tau
\bm{B}_{l(r)}\cdot \bm{\sigma}\right\}}$. After integrating out
the fermion fields, we perform the following gauge transformation:
\begin{eqnarray}
 h_{l(r)}(\tau) &\rightarrow& U_{l(r)}^{\dagger}(\tau)h_{l(r)} \ ,
                \nonumber \\
 h_{l(r)}^{\dagger}(\tau) &\rightarrow&  h_{l(r)}^{\dagger}(\tau)U_{l(r)}
                (\tau) \ . \label{zee3}
\end{eqnarray}
Then, the only effect of the external magnetic fields on the
effective action of the tunnel junction is that the bulk action
[Eq. (\ref{lact1})] turns into
\begin{equation}
 {\mathcal A}_l={\mathcal A}_l^0+M_l\int^{\beta}_0 \! \! d\tau ~\bm{n}_l
               \cdot {\mathrm tr}\left[\bm{\sigma}h_l^{\dagger}\left(
               \partial_{\tau}-\frac{i}{2}~\bm{B}_l\cdot \bm{\sigma}\right)
               h_l\right] \ , \label{lact2}
\end{equation}
and a similar expression with $l\rightarrow r$, and ${\mathcal
A}_T$ [Eq. (\ref{tact2})] remains intact. Eq. (\ref{lact2}) pins
the value of $h_l$ and gives $h_l(\tau)=U_l(\tau)$, and a similar
expression with $l\rightarrow r$. Inserting this into Eq.
(\ref{spinc4}), we find that the last two terms in
$\bm{I}[h_l,h_r;\tau]$ [Eq. (\ref{c1})] vanish after taking the
trace, whereas the first term gives a time-independent component
to the spin current. By choosing $\bm{B}_l=\bm{B}_r\equiv \bm{B}$,
the latter also vanishes. Hereafter, for simplicity, we shall
focus on this situation. (We restrict ourselves to the case where
the gyromagnetic ratio and the effective mass of electrons are
identical for the left and right FM metals.) Under this condition,
the spin current arises solely from the second term in
$\bm{I}[h_l,h_r;\tau]$ and the result is
\begin{eqnarray}
 \langle \bm{I}(\tau)\rangle /I_0 &=& \cos{(B\tau )}\left\{(\bm{n}_l\times
         \bm{n}_r)-\bm{e}\left[\bm{e}\cdot (\bm{n}_l\times \bm{n}_r)\right]
         \right\} \nonumber \\
         & & -\sin{(B\tau )}\left[\bm{e}\times (\bm{n}_l\times \bm{n}_r)
         \right] \nonumber \\
         & & +\bm{e}\left[\bm{e}\cdot (\bm{n}_l\times \bm{n}_r)\right] \ ,
         \label{ac1}
\end{eqnarray}
where $\bm{B}=B\bm{e}$ and $B=|\bm{B}|$.

To understand the meaning of the various terms in Eq. (\ref{ac1}),
we first note that the external magnetic fields will induce a spin
current on account of the precession of the magnetization around
the axis of the magnetic field, which takes the form
$\bm{I}(\tau)\propto \sin{(|\bm{B}|\tau)}\left[\bm{n}-(\bm{n}\cdot
\bm{e})\bm{e}\right]+\cos{(|\bm{B}|\tau)}\left(\bm{e}\times
\bm{n}\right)$ where $\bm{n}$ and $\bm{e}$ are the unit vectors
along the directions of the magnetization and the magnetic field,
respectively. Compared with Eq. (\ref{ac1}), one may recognize
that the first two terms in Eq. (\ref{ac1}) arise from the
precession of $\bm{n}_l\times \bm{n}_r$ around the axis of the
external magnetic field, while the last term is the component of
the DC spin Josephson current parallel to the direction of
magnetic fields, which does not perform the precession. As a
consequence, we will identify the spin current given by Eq.
(\ref{ac1}) as the AC spin Josephson current. To sum up, we have
shown that a {\it time-independent uniform magnetic field} can
induce the AC spin Josephson effect.

\paragraph{Current noise}

Finally, we would like to compute the noise spectrum of the spin
current, which can be extracted from the the two-point correlation
function of the spin current. According to Eq. (\ref{noise1}), the
latter is given by
\begin{widetext}
\begin{eqnarray}
 \langle I_a(\tau_1)I_b(\tau_2)\rangle &=& \frac{1}{{\mathcal Z}[0]}
         \int \! \! D[u]~
         e^{-{\mathcal A}_{eff}}\left\{I_a[h_l,h_r;\tau_1]I_b
         [h_l,h_r;\tau_2]-\frac{1}{4}~K_{ab}[h_l,h_r;\tau_1,\tau_2]
         \right\} \ , \label{noise2}
\end{eqnarray}
where $D[u]=D[h^{\dagger}_l]D[h_l]D[h^{\dagger}_r]D[h_r]$, and
\begin{eqnarray*}
 K_{ab}[h_l,h_r;\tau_1,\tau_2] &=& {\mathcal I}_1(\tau_1-\tau_2)
       \left\{{\mathrm tr}\left[h^{\dagger}_r(\tau_1)\sigma_ah_l
       (\tau_1)h^{\dagger}_l(\tau_2)\sigma_bh_r(\tau_2)\right]
       +(a\leftrightarrow b, \tau_1\leftrightarrow \tau_2)\right\}
       \\
       & & +{\mathcal I}_2(\tau_1-\tau_2)\left\{{\mathrm tr}\left[
       (\bm{n}_r\cdot \bm{\sigma})h^{\dagger}_r(\tau_1)\sigma_ah_l
       (\tau_1)(\bm{n}_l\cdot \bm{\sigma})h^{\dagger}_l(\tau_2)
       \sigma_bh_r(\tau_2)\right]+(a\leftrightarrow b, \tau_1
       \leftrightarrow \tau_2)\right\} \\
       & & +{\mathcal I}_3(\tau_1-\tau_2)\left\{{\mathrm tr}\left[
       h^{\dagger}_r(\tau_1)\sigma_ah_l(\tau_1)(\bm{n}_l\cdot
       \bm{\sigma})h^{\dagger}_l(\tau_2)\sigma_bh_r(\tau_2)\right]
       +(a\leftrightarrow b, \tau_1\leftrightarrow \tau_2)\right\}
       \\
       & & +{\mathcal I}_4(\tau_1-\tau_2)\left\{{\mathrm tr}\left[
       (\bm{n}_r\cdot \bm{\sigma})h^{\dagger}_r(\tau_1)\sigma_ah_l
       (\tau_1)h^{\dagger}_l(\tau_2)\sigma_bh_r(\tau_2)\right]
       +(a\leftrightarrow b, \tau_1\leftrightarrow \tau_2)\right\}
       \ .
\end{eqnarray*}
\end{widetext}
In terms of the semiclassical approximation, the connected
two-point correlation function of the spin current is given by
\begin{eqnarray}
 D_{ab}(\tau_1-\tau_2) &\equiv& -\left[\langle I_a(\tau_1)I_b(\tau_2)
                       \rangle -\langle I_a\rangle \langle I_b\rangle
                       \right] \label{noise3} \\
                       &=& \delta_{ab}{\mathcal I}_1(\tau_1-\tau_2)
                       +R_{ab}{\mathcal I}_2(\tau_1-\tau_2) \ ,
                       \nonumber
\end{eqnarray}
where $R_{ab}=n_{la}n_{rb}+n_{lb}n_{ra}-\delta_{ab}(\bm{n}_l\cdot
\bm{n}_r)$. Note that the third and forth terms in ${\mathcal
A}_T$ only contribute to the higher order correlation functions of
the spin current within the semiclassical approximation.

The noise spectrum is determined by the Fourier transform of the
autocorrelation function of the spin current,
\begin{eqnarray}
 S_{ab}(\omega) &\equiv& \int^{\infty}_{-\infty} \! \! dt~
       e^{i\omega t}\frac{1}{2}\left\langle \left[I_a(t),I_b(0)\right]_+
       \right\rangle \label{noise4} \\
       &=& \frac{1}{2}~\rho_{ab}(\omega)\coth{\left(\frac{\omega}{2T}
       \right)} \ , \nonumber
\end{eqnarray}
where $[,]_+$ denotes the anticommutator. In the above,
$\rho_{ab}(\omega)$ is the spectral function of the
current-current correlation function, which is related to the
Fourier transform of $D_{ab}(\tau)$ via
\begin{equation}
 \rho_{ab}(\omega)=-2{\mathrm Im}\left\{{\mathcal D}_{ab}(\omega +i0^+)
                  \right\} \ , \label{dos1}
\end{equation}
where ${\mathcal D}_{ab}(i\omega_n)=\int^{\beta}_0 \! \! d\tau
~e^{i\omega_n\tau}D_{ab}(\tau)$ and ${\mathcal D}_{ab}(\omega
+i0^+)$ is obtained from ${\mathcal D}_{ab}(i\omega_n)$ through
the analytical continuation $i\omega_n \rightarrow \omega +i0^+$.
From Eq. (\ref{noise3}), we see that to get $\rho_{ab}(\omega)$,
we need the Fourier transforms of ${\mathcal I}_1(\tau)$ and
${\mathcal I}_2(\tau)$ which are, respectively, given by
\begin{eqnarray}
 {\mathcal J}_1(i\omega_n) &=& \frac{|\widetilde{T}|^2}{8}\int \! \!
               d\epsilon_1d\epsilon_2~\frac{N_l(\epsilon_1)N_r
               (\epsilon_2)}{i\omega_n-\epsilon_1+\epsilon_2}
               \nonumber \\
               & & \times \left[\tanh{\left(\frac{\epsilon_1}{2T}\right)}
               -\tanh{\left(\frac{\epsilon_2}{2T}\right)}\right] \ ,
               \nonumber \\
 {\mathcal J}_2(i\omega_n) &=& \frac{|\widetilde{T}|^2}{2}\int \! \!
               d\epsilon_1d\epsilon_2~\frac{n_l(\epsilon_1)n_r
               (\epsilon_2)}{i\omega_n-\epsilon_1+\epsilon_2}
               \nonumber \\
               & & \times \left[\tanh{\left(\frac{\epsilon_1}{2T}\right)}
               -\tanh{\left(\frac{\epsilon_2}{2T}\right)}\right] \ ,
               \label{dos2}
\end{eqnarray}
where ${\mathcal J}_l(i\omega_n)=\int^{\beta}_0 \! \! d\tau
~e^{i\omega_n\tau}{\mathcal I}_l(\tau)$ with $l=1,2$. At low
temperature, i.e. $T\ll \epsilon_F$, the densities of states can
be approximated as the ones at the Fermi surface, and we obtain
${\mathrm Im}\left\{{\mathcal J}_l(\omega +i0^+)\right\}=-\pi
\omega \alpha_l$ with $l=1,2$. Inserting this into Eq.
(\ref{noise4}) gives
\begin{equation}
 S_{ab}(\omega)=(\delta_{ab}\alpha_1+R_{ab}\alpha_2)\pi \omega
               \coth{\left(\frac{\omega}{2T}\right)} \ .
               \label{noise5}
\end{equation}
We see that the noise spectrum of the spin Josephson current takes
the Johnson-Nyquist form at low temperature. This result is not
surprising because the Johnson-Nyquist form is a direct
consequence of a linear circuit, and in our calculations the
semiclassical configuration dominates the path (functional)
integral. If there is some tunable structure in the junction or
the quantum fluctuations are strong, then a deviation from the
Johnson-Nyquist form should be expected.

\section{Renormalization group analysis}
\label{rga}

The underlying assumption behind the semiclassical approximation
is that ${\mathcal A}_T$ in Eq. (\ref{jact4}) is not relevant
under the RG flow, and thus we can treat it as a perturbation in
the weak tunneling limit. This has to be verified by a
renormalization group (RG) analysis.

At zero temperature, ${\mathcal I}_i(\tau)$ with $i=1,\cdots ,4$
are of the form
\begin{equation}
 {\mathcal I}_i(\tau)\approx -\frac{\alpha_i}{\tau^2} \ , ~~~~
                     i=1,\cdots ,4  \ . \label{tact3}
\end{equation}
Therefore, ${\mathcal A}_T$ is a non-local action in the time
space. To proceed, we define the following operators
\begin{widetext}
\begin{eqnarray}
 {\mathcal Q}_1({\mathcal T}) &=& \int \! \! d\tau~\frac{1}{\tau^2}~
               {\mathrm tr}\left[{\mathcal M}^{\dagger}({\mathcal T}
               +\tau /2){\mathcal M}({\mathcal T}-\tau /2)\right] \ ,
               \label{op1} \\
 {\mathcal Q}_2({\mathcal T}) &=& \int \! \! d\tau~\frac{1}{\tau^2}~
               {\mathrm tr}\left[(\bm{n}_r\cdot \bm{\sigma})
               {\mathcal M}^{\dagger}({\mathcal T}+\tau /2)(\bm{n}_l
               \cdot \bm{\sigma}){\mathcal M}({\mathcal T}-\tau /2)
               \right] \ , \label{op2} \\
 {\mathcal Q}_3({\mathcal T}) &=& \int \! \! d\tau~\frac{1}{\tau^2}~
               {\mathrm tr}\left[{\mathcal M}^{\dagger}({\mathcal T}
               +\tau /2)(\bm{n}_l\cdot \bm{\sigma}){\mathcal M}
               ({\mathcal T}-\tau /2)\right] \ , \label{op3} \\
 {\mathcal Q}_4({\mathcal T}) &=& \int \! \! d\tau~\frac{1}{\tau^2}~
               {\mathrm tr}\left[(\bm{n}_r\cdot \bm{\sigma})
               {\mathcal M}^{\dagger}({\mathcal T}+\tau /2){\mathcal M}
               ({\mathcal T}-\tau /2)\right] \ , \label{op4}
\end{eqnarray}
and then, ${\mathcal A}_T$ can be written as
\begin{equation}
 {\mathcal A}_T=-\int \! \! d{\mathcal T}\left[\alpha_1{\mathcal Q}_1
               ({\mathcal T})+\alpha_2{\mathcal Q}_2({\mathcal T})+\alpha_3
               {\mathcal Q}_3({\mathcal T})+\alpha_4{\mathcal Q}_4({\mathcal T})
               \right] \ . \label{op5}
\end{equation}
\end{widetext}
The relevancy of ${\mathcal A}_T$ is determined by the scaling
dimensions of these operators at the fixed point described by the
action
\begin{equation}
 {\mathcal A}_0=\int^{\beta}_0 \! \! d\tau \left\{M_l\bm{n}_l\cdot
               {\mathrm tr}\left[\bm{\sigma}h_l^{\dagger}\partial_{\tau}
               h_l\right]+(l\rightarrow r)\right\} \ . \label{fp1}
\end{equation}
The scaling dimensions of the operators ${\mathcal Q}_i$'s can be
extracted from the long-time behaviors of their two-point
correlation functions $\left \langle {\mathcal Q}_i({\mathcal
T}_1){\mathcal Q}_i({\mathcal T}_2)\right \rangle$. At the fixed
point, they are given by
\begin{equation}
 \left \langle {\mathcal Q}_i({\mathcal T}_1){\mathcal Q}_i({\mathcal T}_2)
 \right \rangle \sim \frac{1}{\left({\mathcal T}_1-{\mathcal T}_2\right)^{2+4d_M}}
        ~~ i=1,\cdots ,4 \ , \label{rg1}
\end{equation}
where $d_M$ is the scaling dimension of ${\mathcal M}$. As a
result, the scaling dimensions of the operators ${\mathcal Q}_i$'s
are identical and are given by $d_Q=1+2d_M$. By the definition of
${\mathcal M}$, we have $d_M=d_l+d_r$ where $d_{l(r)}$ is the
scaling dimension of $h_{l(r)}$. From Eq. (\ref{fp1}), we get
$d_l=0=d_r$. Therefore, $d_Q=1$, and we conclude that all terms in
${\mathcal A}_T$ are marginal perturbations with respect to
${\mathcal A}_0$.

The other way to study the effects of ${\mathcal A}_T$ is to
compute its correction to the free energy.\cite{C} It is given by
\begin{equation}
 \Omega =\Omega_0+\Omega_1+\Omega_2+\cdots \ , \label{cfe1}
\end{equation}
where $\Omega_0$ is the unperturbed free energy, and
\begin{eqnarray}
 \Omega_1 &=& \frac{1}{\beta}\left\langle {\mathcal A}_T\right\rangle \ ,
          \label{cfe2} \\
 \Omega_2 &=& -\frac{1}{2\beta}\left\langle {\mathcal A}_T{\mathcal A}_T
          \right\rangle \ . \label{cfe3}
\end{eqnarray}
In Eq. (\ref{cfe1}), $\cdots$ contains the higher order
correlation functions of ${\mathcal A}_T$. At low temperature, we
have
\begin{eqnarray}
 \Omega_1 &=& -\sum_{i=1}^4\alpha_i\left\langle {\mathcal Q}_i
          \right\rangle \ , \label{cfe4} \\
 \Omega_2 &=& -\frac{1}{2}\sum_{i,j=1}^4\alpha_i\alpha_j\int \! \!
          d\tau \left\langle {\mathcal Q}_i(\tau){\mathcal Q}_j
          (0)\right\rangle \ . \label{cfe5}
\end{eqnarray}
In the above, the expectation values are evaluated at the fixed
point. Because $d_M=0$, one may find that
\begin{eqnarray*}
 \left\langle {\mathcal Q}_i\right\rangle \sim \int \! \! d\tau~
      \frac{1}{\tau^2}\sim \xi_{\tau}^{-1} \ ,
\end{eqnarray*}
as the correlation time $\xi_{\tau}\rightarrow \infty$. Since the
singular part of the unperturbed free energy behaves like
$\xi_{\tau}^{-1}$, we get $\Omega_1/\Omega_0=O(1)$. This implies
that ${\mathcal A}_T$ is a marginal perturbation at the tree
level. The next-order corrections to the RG flow arise from
$\Omega_2$. Because
\begin{equation}
 \left\langle {\mathcal Q}_i(\tau){\mathcal Q}_j(0)\right\rangle =
      \frac{C_{ij}}{\tau^2} \ , \label{rg2}
\end{equation}
where $C_{ij}$ is a numerical constant, we also obtain
$\Omega_2\sim \xi_{\tau}^{-1}$, and thus $\Omega_2/\Omega_0=O(1)$.
In other words, we verify that, to the second order in $\alpha_i$,
${\mathcal A}_T$ is a marginal perturbation. Thus, in the weak
tunneling limit, the use of the semiclassical approximation to
compute the spin current and its correlation functions is
justified. Moreover, in contrast to the SC junctions, the
quasiparticle tunneling does not destroy the quantum coherence
between the two ferromagnets, at least in the weak tunneling
limit.

\section{Conclusions and discussions}
\label{con}

In this paper, we have studied the Josephson-like tunneling
currents between two FM leads within the framework of Stoner
ferromagnetism. In comparison with the previous work, our analysis
maintains the full spin SU($2$) symmetry in all intermediate
steps. More importantly, we clarify the origin of the AC spin
Josephson effect by utilizing an SU($2$) gauge transformation to
probe the nonabelian phase in calculating the AC tunneling spin
current. This approach also reveals most clearly the Josephson
current as a consequence of spontaneous symmetry breaking.

To obtain the AC spin Josephson current [Eq. (\ref{ac1})], a major
simplification we made is that the orbital effects of electrons in
the external magnetic fields are neglected. One way to bypass the
possible complications due to the coupling between the orbital
motion and magnetic fields is to use the FM thin films with an
in-plane magnetic field. In that case, the orbital motion in the
direction perpendicular to the films will be quenched and its
omission is justified.

The explicit form of the AC spin current we obtained [Eq.
(\ref{ac1})] should also have important experimental implications:
It has been suggested that the spin current without charge
currents will induce an electric dipole field.\cite{ML,SKK,SGW}
Therefore, the measurement of the induced electric field can be
used as a detection of the spin Josephson current. In the same
way, the AC spin Josephson current we predicted will induce a
time-dependent electric field with a period $2\pi /(g\mu_BB)$
where $B$ is the magnitude of the external magnetic field. As a
consequence, the detection of a {\it time-dependent} electric
field with the above period in an applied {\it time-independent}
uniform magnetic field may provide a convincing evidence of the AC
spin Josephson effect.

The importance and advantage of the effective action approach also
reveal themselves in clarifying the role of the quasi-particle
tunneling played in the dissipation. In the case of the SC
junctions, in addition to the bulk sector, the effective action
consists of another two terms --- a non-local term due to the
quasiparticle tunneling and a local one arising from the
Cooper-pair tunneling.\cite{AES} The latter is the origin of the
Josephson effect. In that case, the RG analysis indicates that a
strong coupling fixed point exists, where the Cooper-pair
tunneling term becomes irrelevant.\cite{FZ} This result has been
interpreted as the suppression of the Josephson current or the
destruction of the quantum coherence between two superconductors
by the quasi-particle tunneling. As we emphasized in Sec.
\ref{effact}, the most important distinction between the SC and FM
junctions lies in the lack of a pair condensate for the
ferromagnets. Consequently, the effective action for the FM
junction is scale invariant up to the second-order perturbative RG
analysis, where no possible IR instability was found in the
perturbation theory. This implies that the spin Josephson effects
we obtained are robust against the quasi-particle tunneling, at
least for the weak tunneling junctions.

\acknowledgments

Y.L. Lee would like to thank J.C. Wu and Z.H. Wei for discussions.
Y.-W. Lee is grateful to M.F. Yang for discussions. The work of
Y.L. Lee is supported by the National Science Council of Taiwan
under grant NSC 92-2112-M-018-009. The work of Y.-W. Lee is
supported by the National Science Council of Taiwan under grant
NSC 92-2112-M-029-008.


\end{document}